\documentclass[manuscript]{aastex}
\usepackage{natbib}
\usepackage{color}
\usepackage{times}
\usepackage{graphicx}
\usepackage{float}
\usepackage{geometry}
\usepackage{lscape}
\usepackage{rotating}

\newcommand{\ui}{$\pm$}

\shorttitle{Coalescense rate estimation}
\shortauthors{Petrillo et al.}

\begin{document}

\title{Compact object coalescence rate estimation from short gamma-ray burst observations}

\author{Carlo Enrico Petrillo}
\affil{Dipartimento di Scienze Fisiche, Universit\`a di Napoli ``Federico II,''\\ Compl.\ Univ.
Monte S.\ Angelo, Ed.\ N, Via Cinthia, I-80126 Napoli, Italy}

\author{Alexander Dietz}
\affil{Department of Physics and Astronomy, The University of Mississippi\\ University, MS
38677-1848 USA}

\author{Marco Cavagli\`a}
\affil{Department of Physics and Astronomy, The University of Mississippi\\ University, MS
38677-1848, USA}

\begin{abstract} 
Recent observational and theoretical results suggest that Short-duration Gamma-Ray Bursts (SGRBs)
are originated by the merger of compact binary systems of two neutron stars or a neutron star and a
black hole. The observation of SGRBs with known redshifts allows astronomers to infer the merger
rate of these systems in the local universe. We use data from the SWIFT satellite to estimate this
rate to be in the range $\sim 500$-$1500$ Gpc$^{-3}$yr$^{-1}$. This result is consistent with
earlier published results which were obtained through alternative approaches. We estimate the
number of coincident observations of gravitational-wave signals with SGRBs in the advanced
gravitational-wave detector era.  By assuming that all SGRBs are created by neutron star-neutron
star (neutron star-black hole) mergers, we estimate the expected rate of coincident observations to
be in the range $\simeq 0.2$ to $1$ ($\simeq 1$ to $3$) yr$^{-1}$.
\end{abstract}


\maketitle

\section{Introduction}
Short-duration Gamma-Ray Bursts (SGRBs) are some of the most powerful explosions detected in the
universe, releasing intensive bursts of high-energy gamma rays with a peak duration shorter than
two seconds \citep{ck93}. The most commonly accepted explanation of their origin is a system of two
compact objects, either two Neutron Stars (NS-NS) or a Neutron Star and a Black Hole (NS-BH)
coalescing into a black hole \citep{schramm89, Narayan:1992iy, Piro:2005nature, Rezzolla:2011da}.
Because of the emission of Gravitational Waves (GWs) during the latest phases of the binary
evolution, these objects are one of the primary sources for the next generation of ground-based GW
detectors such as Advanced LIGO~\citep{Smith:2009bx} and Advanced Virgo~\citep{Acernese:2008zzf}.
Direct detection of a GW signal from a compact binary coalescing system would allow astronomers to
gain valuable information on the astrophysics of compact objects, for example the NS equation of
state \citep{flanagan:021502, Read:2009}, as well as probe fundamental physics by testing the
Lorentz invariance principle \citep{Ellis2006402} and general relativity in the strong-field regime
\citep{Will:2005va}, or by setting limits on the graviton mass \citep{PhysRevD.80.044002,
Keppel:2010qu}. Direct detection of a GW signal in coincidence with a GRB optical counterpart could
provide additional insights on astrophysics and even cosmology. The measure of the redshift of a
GRB in coincidence with a GW detection could allow astronomers to directly determine the distance
of the system (see, e.g., \citet{Nissanke:2009kt} and references therein). Coincident detections
could also significantly improve the determination of the Hubble parameter by GW observations
\citep{Dalal:2006qt, DelPozzo:2011yh}.

In this context, it is crucial to have reliable estimates of the local merger rate of compact
objects and the number of expected coincident GW-SGRB observations in the advanced GW detector era.
In this paper we present a simple estimate of these quantities by using SGRB data from SWIFT
observations \footnote{\tt http://heasarc.gsfc.nasa.gov/docs/swift/swiftsc.html}. In order to avoid
selective bias, we calculate the number of expected coincident observations by restricting the
sample of SWIFT data to observations with determined redshift and certain association to an optical
counterpart.  In contrast to a previous study by one of the authors \citep{Dietz:2010eh}, we also
include the observed GRB luminosities in the analysis.

The paper is organized as follows. In Section \ref{sec:dataModels} we define the SWIFT data sample
and discuss the theoretical model which is used to fit the observations. In Section
\ref{sec:results} we present the results and compare them to other published rate estimates. The
Appendix contains details on the rate functions used in the analysis.

\section{SWIFT data sample and fitting model}\label{sec:dataModels}
We restrict our analysis to a set of SGRBs with reliable redshift measurement, i.e., to SGRBs that
can be associated to a galaxy of known spectroscopic redshift with a high probability of being the
host galaxy of the SGRB. We omit from the analysis SGRBs without an observed optical afterglow and
SGRBs that are not associated with a host galaxy within the error-circle of the observation. This
allows us to remove any instrumental bias with respect to SGRBs detected by other missions.
Table~\ref{tab:data} shows the list of the 14 SWIFT SGRBs that pass our selection criteria.  

\begin{table*}[th!]
\caption{\label{tab:data} List of the 14 SGRBs observed by SWIFT between 2004 and 2011 which pass
our selection criteria. The table shows the observed fluences, the redshifts and the methods used
to estimate the spectroscopic redshifts of the host galaxies. The fluence data are taken from
\citet{Sakamoto:2007yi} (first 7 SGBRs) and the Gamma-Ray burst Coordinated Network [{\tt
http://gcn.gsfc.nasa.gov/gcn3\_archive.html}] (last 7 SGBRs: \citet{2007GCN..6623....1B},
\citet{2007GCN..7148....1S}, \citet{2008GCN..8187....1C}, \citet{2009GCN..9337....1U},
\citet{2010GCN..10338...1M}, \citet{2010GCN..11111...1M}, \citet{2010GCN..11467...1K},
respectively). The unusual durations of the SGRB 061006 and 070714B are due to light curves with a
short initial event followed by a softer extended event. They are classified as SGRBs (see
\citet{2006GCN..5699....1S} and \citet{2007GCN..6623....1B}). All SGRBs are preceded by an
afterglow except 060502B, 060801 and 101219, however only one galaxy was present in the error
circle for these SGRBs.}
\scriptsize
\begin{center}
\begin{tabular}{|l|c|c|c|c|c|c|}
\hline 
 GRB &  Duration & $z$ & type & Reference  & Fluence                     \\
     &        [s]  &  {} & {} &    {} & {[}$10^{-7}$erg/cm$^2${]}    \\
\hline  
\hline 
 050416  &    2.0  & 0.6535 &  emission  &  \citet{2005GCN..3542....1C}, \citet{2007ApJ...661..982S} &  3.7\ui 0.4 \\
\hline 
 051221  &    1.4  & 0.5465 & emission  & \citet{2005GCN..4384....1B}, \citet{2006ApJ...650..261S} & 11.5\ui 0.4 \\
\hline 
 060502B &    0.09 & 0.287  & absorption & \citet{2006GCN..5238....1B},  \citet{2007ApJ...654..878B} &  0.4\ui 0.1 \\
\hline 
 060801  &    0.5  & 1.131  & emission  &  \citet{Cucchiara2006},  \citet{Berger:2006ik} &  0.8\ui 0.1 \\
\hline 
 061006  &  130    & 0.4377 & emission  & \citet{Berger:2006ik} & 14.2\ui 1.4 \\
\hline 
 061201  &    0.8  & 0.111  & emission  & \citet{2006GCN..5952....1B}, \citet{2007AA...474..827S} &  3.3\ui 0.3 \\
\hline
 070429B &    0.5  & 0.9023 & emission  &  \citet{2007GCN..7140....1P},  \citet{2008arXiv0802.0874C}  &  0.6\ui 0.1 \\
\hline 
 070714B &   64    & 0.9225 & emission  & \citet{2007GCN..6836....1G}, \citet{2008arXiv0802.0874C} &  5.1\ui 0.3 \\
\hline 
 071227  &    1.8  & 0.384  & emission  & \citet{2007GCN..7154....1B},  \citet{2009AA...498..711D} &  2.2\ui 0.3 \\
\hline 
 080905  &    1.0  & 0.1218 & emission  & \citet{Rowlinson:2010jb}&  1.4\ui 0.2 \\
\hline 
 090510  &    0.3  & 0.903  & emission  & \citet{2009GCN..9353....1R}, \citet{2010AA...516A..71M} &  3.4\ui 0.4 \\
\hline 
 100117  &    0.3  & 0.92   & emission  & \citet{Fong:2010kz} &  0.9\ui 0.1 \\
\hline 
 100816  &    2.9  & 0.8049 & absorption& \citet{2010GCN..11125...1G} & 20.0\ui 1.0  \\
\hline
 101219  &    0.6  & 0.718  & emission  & \citet{2011GCN..11518...1C} &  4.6\ui 0.3 \\
\hline 
\end{tabular}
\end{center}
\end{table*}

\begin{figure}[t!]
\begin{center}
\includegraphics[width=3.25in,angle=0]{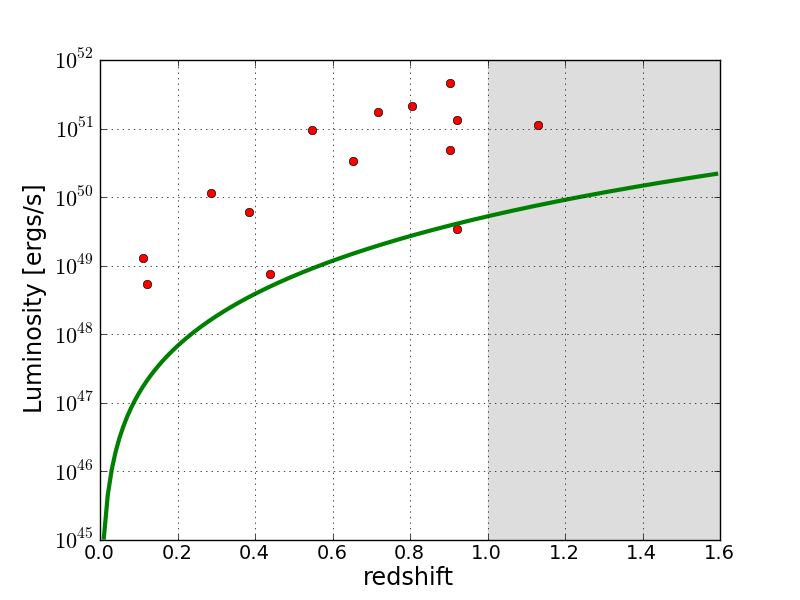}~
\includegraphics[width=3.25in,angle=0]{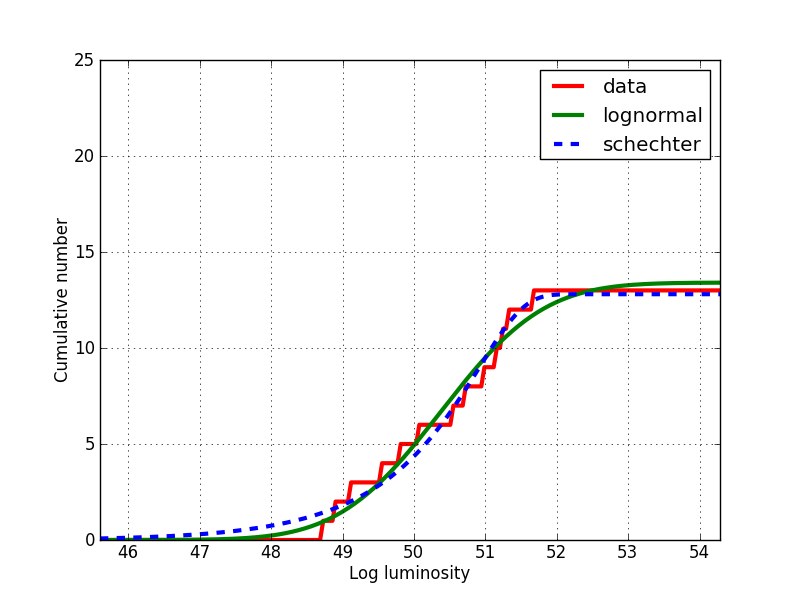}
\end{center}
\caption{Left panel: Luminosity distribution of the 14 SGRBs which pass the analysis selection
criteria as function of their redshift. The grey area represents the \textit{redshift desert}. Our
analysis is restricted to the 13 SGRBs with $z<1$. Right panel: Cumulative distribution of the
SGRBs as a function of luminosity (red), the lognormal function fit (green), and the Schechter
function fit (dashed blue).}
\label{fig:lumdistr}
\end{figure}

The luminosity of these SGRBs can be computed using their redshift and fluence information. The
fluence $S$ is divided by the SGRB duration to estimate the flux $F$ in the relevant frame for the
detection threshold of the satellite (the observer frame). Since the observed fluence depends on
the spectral properties of the source and the energy response of the detector (15 to 150 keV for
the BAT instrument onboard SWIFT, see \citet{Barthelmy:2005hs}), and the observations are over
cosmological distances, two identical sources at different distances may show a spectral shift and
a change of fluence. Expressing the observed photon number spectrum with the Band function
\citep{Band:1993eg}, this spectral shift can be calculated as a function of the redshift
\citep{Cao:2011ar}. Since most SGRB sources have redshift smaller than $z=$1, the effect of the
spectral shift is less than 10\% \citep{Cao:2011ar}. This error is small compared to other
statistical and systematical errors and will be neglected in our analysis. The apparent luminosity
of the SGRB is
 \begin{equation}
L= 4\pi d_{\mathrm{L}}^{2}(z)\,F \simeq 4\pi d_{\mathrm{L}}^{2}(z)\frac{S}{T_{90}}\,,
\end{equation}
where $d_{\mathrm{L}}(z)$ is the luminosity distance for a given redshift $z$, $F$ is the mean
flux,  $S$ is the measured fluence and $T_{90}$ is the time over which the burst emits 90\% of its
total energy.  Throughout this paper we consider a standard flat-$\Lambda$ cosmology with $H_0=71\;
\mathrm{km}\;\mathrm{s}^{-1}$, $\Omega_M=0.27$ and  $\Omega_\Lambda=0.73$. The left panel of Fig.\
\ref{fig:lumdistr} shows the distribution of the observed luminosities as a function of the
redshift. The solid line indicates the approximate detector's sensitivity threshold
\begin{equation}
d_\mathrm{max}(L) = \sqrt{\frac{L}{4\pi F_\mathrm{lim}}}\,,
\label{eq:threshold}
\end{equation}
where the flux threshold is $F_\mathrm{lim}=5\times 10^{-9}$~erg s$^{-1}$ cm$^{-2}$ 
\citep{Cao:2011ar}. The grey area in Fig.\ \ref{fig:lumdistr} defines the so-called
\textit{redshift desert}, a region between $z\simeq 1$ and $z\simeq 2$ where spectroscopic redshift
determinations are difficult to obtain \citep{Fiore:2007yk}. In the following, we choose a
conservative approach and further restrict our data sample to $z<1$, leaving 13 data points for our
analysis. 

To determine the local merger rate, we fitted the sample with several, commonly-used luminosity
functions \citep{Dietz:2010eh, Chapman:2008zx, Guetta:2008qw}. The lognormal function and the
Schechter function were found to provide acceptable fits.  The cumulative distribution of the SGRBs
as a function of the luminosity and the lognormal and Schechter functions are shown in Fig.\
\ref{fig:lumdistr}. Since the lognormal fit is slightly better than the Schechter fit, we restrict
our analysis to the lognormal luminosity function
\begin{equation}
\phi(L)\propto \frac{1}{L}\mathrm{exp}\left(-\frac{\mathrm{log}L-\mathrm{log}L_{0}}{2\sigma^{2}}\right)\,,
\label{eq:lumfunction}
\end{equation}
where $L_0$ is the mean (peak) value of the luminosity and $\sigma$ is the width of the
distribution. These two parameters are determined by fitting the function to the 13 SGRBs in the
sample. For sake of simplicity, we do not consider evolutionary effects on the luminosity function.
Although it is reasonable to assume that these effects are small compared to other statistical and
systematical errors, some physical processes depend on the metallicity of the progenitors, which is
a function of the redshift \citep{Belczynski:2010tb, Belczynski:2011qp}. The right panel of Fig.\
\ref{fig:lumdistr} shows the cumulative distribution of SGRBs as a function of luminosity and the
lognormal fit in Eq.\ (\ref{eq:lumfunction}). 

\begin{figure}[t!]
\begin{center}
\includegraphics[width=3.5in,angle=0]{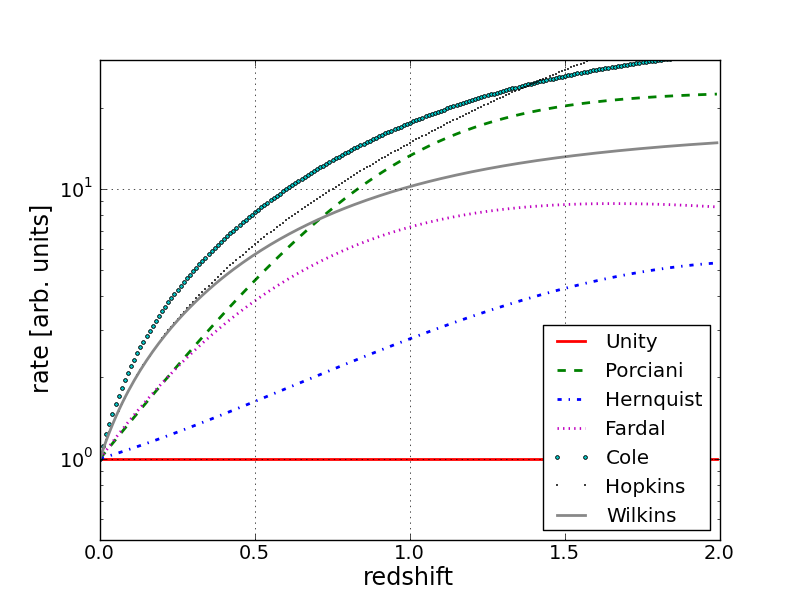}
\caption{Different rate functions used in the analysis (see the Appendix for details). Note that the rate functions vary significantly even for redshift distances $z<1$.}
\end{center}
\label{fig:ratefunctions}
\end{figure}

Since the fit is made on \textit{observed} SGRBs, the luminosity function $\phi'(L)$ must be
rescaled to the volume where SWIFT is sensitive. Assuming an isotropic distribution of SGRBs, we write
\begin{equation}
\phi(L) \propto \phi'(L)/ d_{\mathrm{max}}^{3}(L)\,,
\label{eq:lumcorrection}
\end{equation}
where $d_{\mathrm{max}}^{3}(L)$ is the maximum luminosity distance where a SGRB can be detected by SWIFT. The number of observable SGRBs within a redshift distance $z$ is
\begin{equation}
N'(z)=N_{0}\int_{0}^{z}\mathrm{d}z'\frac{R(z')}{1+z'}\frac{\mathrm{d}V(z')}{\mathbf{\mathrm{d}}z'}\int_{L_{\mathrm{min}}(z')}^{\infty}\phi(L)\mathrm{d}L\,,
\label{eq:mainnumber}
\end{equation}
where $\mathrm{d}V(z')/\mathbf{\mathrm{dz'}}$ is the comoving volume element and $N_{0}$ is a
normalization factor. The rate function $R(z)$ describes the formation rate of the binary systems
per comoving volume as a function of the redshift. Since a binary system of compact objects is
formed from massive progenitor stars, $R(z)$ may be assumed to follow the star formation rate. The
time difference from the formation of the compact objects to the coalescence of the binary is
likely on the order of the Gyr \citep{Belczynski:2006br}. Thus there is a significant delay with
respect to the star formation rate. This is taken into account by using a delayed rate function in
Eq.\ (\ref{eq:mainnumber}). Figure \ref{fig:ratefunctions} shows different rate functions that are
used in our analysis. The Appendix contains explicit expressions and references. 

Since Eq.\ (\ref{eq:mainnumber}) describes the number of SGRBs that may be potentially observed,
$N'(z)$ must be rescaled to fit the number of SGRBs used in the analysis. From 2005 through 2011
($T\simeq 6$ yr) SWIFT observed 46 SGRBs. (We neglect any downtime of the satellite due to
technical issues or other constraints.) Thus $N'(z)$ must be divided by the factor $f_R=14/46$,
with an estimated error of $\sqrt{14}/46\sim 10$\%. The SWIFT field of view is about 1.4~sr
\citep{Barthelmy:2005hs}, corresponding to a visible fraction of the sky approximately equal to
$f_{FOV}\simeq 10$\%. Observations have shown that about 15\% of all SGRBs may be created by Soft
Gamma Repeaters (SGRs) \citep{Nakar:2005bs, Chapman:2008zx}. Since SGRs are typically less bright
than ordinary SGRBs, events at larger redshifts might be composed mainly of SGRBs. If we assume
that 85\% of all SGRBs are created by the merger of two compact objects, the factor
$f_\mathrm{SGR}=0.85$ yields a conservative limit on the merger rate. In order to create a
relativistic outflow, a torus must be created around the newly formed black hole. Simulation and
theoretical analyses show that the formation of a GRB depends on several parameters such as spin of
the black hole, the mass ratio of the binary or the compactness of the neutron star
\citep{Pannarale:2010vs, Rezzolla:2011da}. Since these parameters are hard to generalize, we simply
assume that all compact object mergers produce a GRB. GRBs are believed to emit their radiation in
a collimated cone. The half-opening angle $\theta$ defines the fraction of the sky where the burst
can be seen, $f_b=1-\cos{\theta}$. The angle $\theta$ is highly uncertain, especially in the case
of SGRBs, as it depends on the model and the Lorentz factor of the outflow
\citep{Panaitescu:2001bx}. Measurements of SGRB half-opening angles range from a few degrees to
over 25$^\circ$ \citep{2004Natur.430..648S, 2006ApJ...653..468B, 2006ApJ...653..462G, Panaitescu:2006, Racusin:2008bx}.
In the following, we set  $1/f_b=15$, corresponding to a half-opening angle $\theta\simeq
20^\circ$  \citep{Bartos:2011jy}. By taking into account all these factors, the merger rate
$R_{\mathrm{merger}}(z)$ and the expected rate of SGRB observations $R_{\mathrm{SGRB}}(z)$ are
\begin{equation}
R_{\mathrm{merger}}(z)=\frac{f_\mathrm{SGR}}{T\,f_b\,f_{FOV}\,f_R} N'(z)
\end{equation}
and
\begin{equation}
\label{eq:sgrbrate}
R_{\mathrm{SGRB}}(z)=\frac{1}{T} N'(z)\,,
\end{equation}
respectively.

\section{Results and Discussion}\label{sec:results}
\begin{figure}[t!]
\begin{center}
\includegraphics[width=3.5in,angle=0]{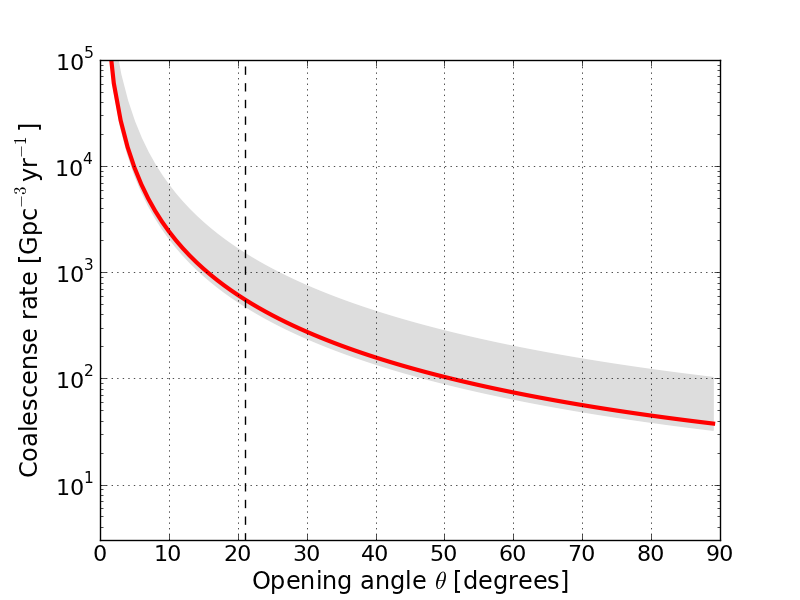}
\caption{Merger rate of compact objects as a function of the half-opening angle $\theta$. The red line indicates the median result. The grey area spans the possible range of rates due to different fit models and systematic errors. The vertical dashed line indicates the typical opening angle which is used in our estimate.}
\label{fig:theta}
\end{center}
\end{figure}

The results of our analysis are summarized in Table \ref{tab:results}. The approximate local merger
rate of binary compact objects ranges from 479 to 1025 Gpc$^{-3}$yr$^{-1}$, depending on the chosen
rate function. The upper limit comes from the model with unity rate function, which assumes no
evolution on star formation over cosmological distances. The Porciani delayed rate function with
delay times of 20 Myr and 100 Myr gives merger rates about 50\% larger than without the delay.
Extrapolating this result to the other functions, a reasonable estimate for the merger rate in the
local universe is in the range $\simeq 500$ to $1500$ Gpc$^{-3}$yr$^{-1}$. These results strongly
depend on the half-opening angle of the SGRB jets. Figure \ref{fig:theta} shows the dependence of
the merger rate on the opening angle $\theta$. The smaller the angle, the larger is the number of
mergers because the observer must be in the outflow cone to detect the SGRB. Assuming an
half-opening angle of 10$^\circ$, the merger rate could be as high as several thousand
Mpc$^{-3}$yr$^{-1}$. A more isotropic large half-opening angle of 60$^\circ$ yields a rate of the
order of 100 Mpc$^{-3}$yr$^{-1}$.

\begin{table}[b!]
\caption{Estimates of merger rates and number of detections per year for Advanced
LIGO/Virgo. The detector reach is 450 Mpc for NS-NS binary systems and 930 Mpc for NS-BH binary systems. Results for different rate functions are shown. The Porciani20 and Porciani100 rate functions include delay times of 20 Myr and 100 Myr, respectively.}
\label{tab:results}
\begin{center}
\begin{tabular}{|c|c|c|c|}
\hline
Rate function & Merger rate & NS-NS Detections & NS-BH Detections \\ 
          & Gpc$^{-3}$yr$^{-1}$  & yr$^{-1}$ & yr$^{-1}$ \\ \hline
            Unity & 1025 &  525 & 3456  \\  
      Hernquist &  816 &  393 & 2750  \\ 
         Fardal &  580 &  238 & 1954  \\ 
           Cole &  485 &  170 & 1634  \\ 
        Hopkins &  506 &  188 & 1706  \\ 
        Wilkins &  553 &  206 & 1866  \\ 
       Porciani &  479 &  196 & 1614  \\
       Porciani20 &  729 &  326 & 2457  \\
       Porciani100 &  757 &  340 & 2552  \\ 
 \hline
\end{tabular}
\end{center}
\end{table}

Assuming that a satellite with a field-of-view comparable to the field of view of SWIFT is
operating at the time of advanced GW detectors, and using the expected range for Advanced
LIGO/Virgo \citep{LVC:2010cfa}, we can estimate the number of coincident observations of SGRBs with
GW counterparts. Under the above assumptions, we estimate about $0.2$ to 1 coincident observations
per year for a NS-NS mergers progenitor (detector range $\simeq 450$ Mpc) and about 1 to 3
coincident observations per year for a NS-BH progenitor (detector range $\simeq 930$ Mpc), a result
consistent with earlier estimates (see \citet{Metzger:2011bv}; \citet{Bartos:2012vd} and references
therein).  These values include the systematic uncertainties underlying the estimates, as explained
above.  Since the advanced detectors are expected to operate for several years, a few observations
of coincident SGRB-GW events seem likely. 

These estimates could improve significantly with a network of operating GRB satellites.  Assuming
that Fermi\footnote{\tt http://fermi.gsfc.nasa.gov/} will be in operation during the advanced
detector era, as well as the planned SVOM mission \citep{Paul:2011ii} and Lobster \footnote{Neil
Gehrels, private communication, 2011}, the coincident SGRB-GW detection rate could be higher than
the above estimate by a factor $\simeq 3$. Ensuring that at least one GRB mission is operational at
the time of advanced detectors will be crucial for identifying the host galaxy, measuring its
redshift and star formation rate, and gaining valuable astrophysical information. 

Our estimates can be compared to earlier published results which were obtained with alternative
approaches. The two most common methods that are used to estimate the merger rate of compact
objects rely on deriving the rates from observed pulsar observations \citep{Kalogera:2003tn} or
employing population synthesis models \citep{Belczynski:2006zi}. Both approaches inherit large
statistical or systematical errors. A recent review article summarizes these results, concluding
that the rate of merger events is somewhere between 10 and 10000 Gpc$^{-3}$yr$^{-1}$
\citep{LVC:2010cfa}. Other investigations rely on methods that are more similar to the method used
here. \citet{Guetta:2005xxx} use a sample of 5 SGRBs to estimate a merger rate in the range 8-30 Gpc~$^{-3}$yr$^{-1}$. An earlier analysis by one of the authors which is based on a less
restrictive data sample and neglects individual GRB luminosities, yields a much higher rate of
about 7800 Gpc$^{-3}$yr$^{-1}$ \citep{Dietz:2010eh}. Finally, a recent study by
\citet{Coward:2012gn} uses a different method based on single GRB observations. In this approach,
the maximum distance at which individual SGRBs can be detected by SWIFT is calculated and the
results are then combined to estimate a final local rate of 0.16 to 1100 Gpc$^{-3}$yr$^{-1}$. These
results show that our estimates are consistent with, and confirm previous merger rate estimates.
Future SGRB observations and improved statistics may further strengthen this conclusion.

\begin{acknowledgements}
This work is the result of a Research Experience for Undergraduates (REU) project by Carlo Enrico
Petrillo at the University of Mississippi. C.E.P., A.D., and M.C.\ are partially
supported by the National Science Foundation through awards PHY-0757937 and PHY-1067985. The
authors would like to thank Jocelyn Read, Emanuele Berti, Maurizio Paolillo, Neil Gehrels and
Richard O'Shaughnessy for their help and valuable comments. This publication has been assigned LIGO
Document Number LIGO-P1200015.  
\end{acknowledgements}

\appendix
\section{Rate functions}
\label{app:rates}
This Appendix describes the various rate functions which are used in the analysis.
\paragraph{Porciani.}
The Porciani rate function is the SF2 function in \citet{Porciani:2000ag}:
\begin{equation}
R(z) \propto \frac{\exp{(3.4\,z)}}{\exp{(3.4\,z)}+22}\,.
\end{equation}
\paragraph{Hernquist.}
The Hernquist rate function is \citep{Hernquist:2002rg}
\begin{equation}
R(z) \propto \frac{\chi^2}{1+\alpha(\chi-1)^3\exp{(\beta \chi^{7/4})}}\,,
\end{equation}
where
\begin{equation}
H(z) = H_0\,\sqrt{(1+z)^3\Omega_M + \Omega_\Lambda}\,,~~~~~~\chi(z) = \left[\frac{H(z)}{H_0}\right]^{2/3}\,,
\end{equation}
and $\alpha=0.012$, $\beta = 0.041$.
\paragraph{Fardal.}
The Fardal rate function is defined as \citep{Fardal:2006sd}
\begin{equation}
R(z) \propto\frac{a^{-p_2}}{(1+p_1\,a^{-p_2})^{p_3+1}} H(z)
\end{equation}
where $p_1=0.075$, $p_2=3.7$, $p_3=0.84$ and $a=(1+z)^{-1}$.
\paragraph{Cole.}
The Cole rate function is 
\begin{equation}
R(z) \propto\frac{a+bz}{1+(z/c)^d}H(z)
\end{equation}
where the parameters are summarized for different authors in Table \ref{tab:coleParams}.
\begin{table}[ht!]
\caption{Parameters of the Cole rate function for different models in the literature.}
\label{tab:coleParams}
\begin{center}
\begin{tabular}{|l|l|l|l|l|}
\hline
Reference & a & b & c & d \\ \hline
Cole \citep{Cole:2000ea} & 0.0166 & 0.1848 & 1.9474 & 2.6316 \\ 
Hopkins \citep{Hopkins:2006bw} & 0.0170 & 0.13 & 3.3 & 5.3 \\ 
Wilkins \citep{Wilkins:2008qc} & 0.014& 0.11& 1.4& 2.2 \\ \hline
\end{tabular}
\end{center}
\end{table}
\paragraph{Delayed functions.}
Since the time between the formation of the compact objects and the merger of the binary system is
typically of the order of the Gyr, the merger rate may not follow directly from the star formation
rate. Assuming a delay with respect to the star formation rate, the delayed rate function is
defined as
\begin{equation}
\label{eq:delay}
R_t(z) = \int_0^{t_d} \mathrm{d}t \frac{1}{1+z_f} R(z_{ret})\,P(t)\,,
\end{equation}
where $P(t)$ represents the probability distribution of the delay time. Population synthesis models \citep{Belczynski:2001uc,Postnov:2007jv} suggest that this distribution is a power law 
\begin{equation}
P(t)\propto t^\alpha\,,
\end{equation}
where $\alpha \simeq -1$ for $t>t_\mathrm{tmin}$. Although the observational literature has applied
a much broader range of functional forms, a time delay probability distribution $P(t)\sim 1/t$ is
sufficient for the purposes of this analysis, where binaries are produced in the
field\footnote{Richard O'Shaughnessy, private communication.}. The retarded redshift, i.e. the
redshift at the time when the compact objects are formed, is
\begin{equation}
\label{eq:retardedz}
z_\mathrm{ret}=T^{-1} \left( T(z)+t \right)\,. 
\end{equation}
The lookback time at redshift $z$ is (see, e.g., \citet{Hogg:1999ad})
\begin{equation}
T(z) = T_0 \int_0^{dz'} \frac{dz'}{(1+z')\sqrt{(\Omega_k(1+z')^3 + \Omega_k(1+z')^2 +  \Omega_\Lambda)}}\,.
\label{eq:lookbacktime} 
\end{equation}
The expression in Eq.\ (\ref{eq:retardedz}) requires an integration and function inversion that in
general need to be evaluated numerically. However, if flat cosmological models with $\Omega_k=0$
are considered, it is possible to obtain an analytic expression for the lookback time. Substituting
\begin{equation}
v = \Omega_k(1+z)^3 
\label{eq:funcv}
\end{equation}
in Eq.\ (\ref{eq:retardedz}), the integral takes the form 
\begin{equation}
T(z) = \frac{T_0}{3} \int_{v(0)}^{v(z)} dv' \frac{1}{v\sqrt{v+\Omega_\Lambda}}\,.
\end{equation}
Integrating, it follows
\begin{equation}
T(z) = \frac{T_0}{3\sqrt{\Omega_\Lambda}} \left(L(v_0)-L(v_1)\right)\,,
\label{eq:funcTz}
\end{equation}
where
\begin{equation}
L(v) = \ln \left( \frac{\sqrt{v+\Omega_\Lambda}+\sqrt{\Omega_\Lambda}}{\sqrt{v+\Omega_\Lambda}-\sqrt{\Omega_\Lambda}}  \right) \,.
\label{eq:funcLv}
\end{equation}
This analytic expression can be used to calculate the lookback time for a given redshift. Solving Eq.\ (\ref{eq:funcTz}) for $L(v_1)$, we find
\begin{equation}
L(v_1)=L(v_0)-3\sqrt{\Omega_\Lambda}\frac{T}{T_H} \equiv \ln{E(T)}\,.
\label{eq:funcLv1}
\end{equation}
The value of $L(v_0)$ does not depends on $z$ or $T$. Equation (\ref{eq:funcLv1}) is a function of
$\Omega_M$ and $\Omega_\Lambda$, as one can see from Eq.\ (\ref{eq:funcLv}) and Eq.\
(\ref{eq:funcv}). Solving Eq.\ (\ref{eq:funcLv}) for $z$, one finally obtains
\begin{equation}
z(T) = \left(\frac{\Omega_\Lambda}{\Omega_M}\right)^{1/3} \left[   \left(\frac{1+E(T)}{1-E(T)}\right)^2  -1\right]^{1/3}-1\,.
\end{equation}
\clearpage
\bibliographystyle{apj}
\bibliography{sgrb-revised-v2}
\end{document}